# Understanding the Use of Voice Assistants by Older Adults

**Margot Hanley**
Cornell Tech
NYC, NY 94022, USA
margothanley@gmail.com

**Shiri Azenkot**
Cornell Tech
NYC, NY 94022, USA
Shiri.azenkot@cornell.edu

## Abstract
Older adults are using voice-based technologies in a variety of different contexts and are uniquely positioned to benefit from smart speakers' hands-free, voice-based interface. In order to better understand the ways in which older adults engage with and learn how to use smart speakers, we conducted qualitative, semi-structured interviews with four older adults who own smart speakers. Emerging findings indicate that older adults benefit from smart speakers as both an assistive and a social technology. Findings also suggest that when older adults learn new technologies in a formal, communal environment there is successful adoption.

## ACM Classification Keywords
H.5.m. Information interfaces and presentation (e.g., HCI): Miscellaneous

## 1. Introduction
Twenty percent of U.S. households own a smart speaker, of which older adults (+65) are a small but rapidly increasing proportion. Smart speakers are an interesting technology for older adults for a few reasons; smart speakers utilize speech input and output, which is a modality of interaction often easier for older adults. Smart speakers are also designed to be sociable; they have names, friendly voices, personalities (they are even programmed to make jokes.) Lastly, there is a very low learning curve to use a smart speaker; they are designed to interact just like another human would, through language.

Older adults are a population that benefits both from social and accessible technology, but there is very little research which explores how older adults are using smart speakers today. What role do the devices play in daily life? What does the learning curve actually look like? This exploratory study focuses on the in situ lived experiences of four older adults living in a retirement home, each of whom own smart speakers. Specifically, this study reveals preliminary themes in the introduction of smart speakers in the lives of older adults.

## 2. Background and Related Work
There has been a significant amount of research conducted on the design and user experience of voice-based systems [3]. Due to the unique ("hands-free") modality of voice-based technology, some researchers have focused specifically on how speech input / output supports accessibility, such as in the case of speech or physical impairments [1].

The voice interface is also an arguably social one [7}. There is been a large body of research which suggests that owners engage in social interactions with smart speakers, they personify them [2] or relate to them as a companion [8]. While there is a lot of research indicating that many older adults experience loneliness and social isolation [6], there hasn't been any research investigating the potential for social interactions between older adults and smart speakers.

While there has been research on how older adults are using and benefitting from voice-based systems, broadly [3], there is no research on how older adults use smart speakers, specifically. Similarly, there is a lot of research focused on how older adults learn new technologies like mobile phones[9], but there is no research focusing on how older adults learn how to navigate interactions with smart speakers.

We situate our research as extensions- and at the intersection of- each of these bodies of work. This study aims to address the gap in the literature by exploring the ways in which older adults are using and benefitting from smart speakers, both from a usability and a social perspective.

## 3. Methods

We conducted 4 semi-structured phone interviews individuals living in a retirement community, each of whom owned a smart speaker. We interviewed two individuals and one married couple. Each interview lasted for about one hour. Our goal with the interviews was to learn about the patterns and challenges for older adults as they use and interact with smart speakers, including the learnability and usability of the device, the role that the device plays in their daily lives, how they deal with challenges, and the social context of its use. We transcribed audio recordings and coded using NVivo. In the following section we discuss the emergent themes and findings.

## 4. Emergent Themes

Below we discuss two large themes which emerged out of interviews; we discuss two roles that smart speakers play in the lives of older adults and how older adults learn to use the speakers.

### 4.1 Smart Speaker Roles

*Assistive Technology*
Each of the four participants brought up how helpful the device was for older adults with disabilities. Herb, a man in his 80's living in the retirement home, mentioned how useful the voice interface was after one of his recent surgeries, *"And I just had cataract surgery last week and you know, you have to have your eye drops three times a day and then you have to wait five minutes between. So, you know, [my wife] can just say, "Alexa, set timer for".* Participants who didn't identify as having a disability instead told stories about fellow residents. Clementine, a woman in her mid-90's, described how a blind, deaf resident benefited from the voice interface design, *"We have a woman here who's a hundred years old, she's partially blind and rather deaf and she has an Echo in her apartment...and she has it read the Bible to her every day and she asks it about the weather and it's been a great comfort to her."*

*Social Interaction*
Each of the four participants talked about ways in which the smart speaker provided a level of connection or companionship or prevented them from feeling "alone". Bessie, a female resident in her 80's, said, *"It's just the idea that you're not alone, you know... I don't come in and welcome Alexa or anything but I mean the practical things such as the information I need every day, it just kind of grows on me. You know, you just got somebody* there. But I don't, I mean it's such a strong point, but it is true that every day I do depend on her." Frannie, another female resident in her 80's, said, *"It also become like a friend, you know, and especially if they read you books or you want to listen to music. You don't feel so... living in silence, I guess."* Herb, her husband, agreed, *"Yeah. And I don't know whether it's a friend or not friend, but you're not alone, even if you, you know, because you can just reach out verbally and get music."*

Clementine pointed out how the older adults in the retirement home who have a physical disability often rely more heavily on the smart speakers for social interactions. She related a story about one of her fellow residents, *"...with her hearing shutdown and her eyesight dimming she spends a lot of time alone. She can't see or hear what's going on, so she's-- it tends-- those physical disabilities tend to isolate you and that's not good for you... Well, sometimes you just can't [socialize] and sometimes you don't feel like it…"* Speaking more broadly, Clementine notes that *"[the smart speaker] is a very useful and fun tool, but if you were all alone in your room, hour after hour, and the only people that came to see you were the nurse to help you get dressed in the morning and the people from the dining room to bring you your meals, if that's the only companionship you had sort of, that and your doctor occasionally dropping by, you would really tend to personalize this more, the device more. It's wonderful for shut ins."*

**4.2 Learnability and Usability of Smart Speakers**

Interviews revealed that each of the four participants had been enrolled in a formal, six-week training offered by Amazon. The course met weekly at the retirement home, and during each session the participants were trained on the smart speaker capabilities. Clementine described the structure of the course, *"Those of us who had the device in our apartments were invited to come to a little get together down at the… classroom here in the building, and the people who were running experiments, the, I say "kids," they're not kids they're in their thirties, from headquarters were there and they would have researched one or two new skills for Alexa and they would tell us how to use them and they would run us through rehearsal and how to use them."*

Each of the four participants emphasized how valuable the formal training was and how easy it made the learning process. Bessie indicated how easy the training made her adoption, *"We had a meeting once a week. I just can't say enough good-- well everyone's mind works differently I guess… But with me, I like to see the picture. I like to have someone tell me what it does and then let me do it. And they go over these things every six weeks to learn a little bit more every week, it's just a very easy."*

Herb talked about how he and fellow residents continued to learn about the smart speakers, collectively. *"It was nice, you know, and they took us through it and you know, we didn't learn everything in the few weeks that we had the class, but then we started our technology group here, we had our own set of classes and then we would talk amongst ourselves what we were learning and, so we came up, you know, kind of a mutual learning process."*

**Conclusion and Future Research**

We know from past research that older adults are a population that benefits both from social and accessible technology. This study, by delving into the behaviors, perceptions, and feelings of four participants, reveals preliminary themes in the introduction of smart speakers in the lives of older adults.

Our preliminary and exploratory analysis indicates that there is potential for these speakers to play a positive, social role in the lives of older adults. The research also supports former research which indicates that that smart speakers- and voice technology more broadly- is enormously beneficial for older adults who have a physical disability.

Future research should focus on increasing the sample size and diversity of subjects. In addition, researchers should focus on furthering our understanding of how older adults benefit from smart speakers' social interface.

**Acknowledgements**
We thank Cornell Tech and the participants.

**References**

[1] Alisha Pradhan, Kanika Mehta, and Leah Findlater. 2018. "Accessibility Came by Accident": Use of Voice-Controlled Intelligent Personal Assistants by People with Disabilities. In Proceedings of the 2018 CHI Conference on Human Factors in Computing Systems (CHI '18). ACM, New York, NY, USA, Paper 459, 13 pages. DOI: https://doi.org/10.1145/3173574.3174033

[2] Amanda Purington, Jessie G. Taft, Shruti Sannon, Natalya N. Bazarova, and Samuel Hardman Taylor. 2017. "Alexa is my new BFF": Social Roles, User Satisfaction, and Personification of the Amazon Echo. In Proceedings of the 2017 CHI Conference Extended Abstracts on Human Factors in Computing Systems (CHI EA '17). ACM, New York, NY, USA, 2853-2859. DOI: https://doi.org/10.1145/3027063.3053246

[3] Aung Pyae and Tapani N. Joelsson. 2018. Investigating the usability and user experiences of voice user interface: a case of Google home smart speaker. In Proceedings of the 20th International Conference on Human-Computer Interaction with Mobile Devices and Services Adjunct (MobileHCI '18). ACM, New York, NY, USA, 127-131. DOI: https://doi.org/10.1145/3236112.3236130

[4] Robin Brewer. 2016. Connecting Older Adults through Voice-Based Interfaces. In Proceedings of the 19th ACM Conference on Computer Supported Cooperative Work and Social Computing Companion (CSCW '16 Companion). ACM, New York, NY, USA, 131-134. DOI: https://doi.org/10.1145/2818052.2874350

[5] Richard J. Wirth, Chien Wen Yuan, Benjamin V. Hanrahan, John M. Carroll, Mary Beth Rosson, and Jomara Bindá. 2016. Exploring Interactive Surface Designs for Eliciting Social Activity from Elderly Adults. In Proceedings of the 2016 ACM International Conference on Interactive Surfaces and Spaces (ISS '16). ACM, New York, NY, USA, 403-408. DOI: https://doi.org/10.1145/2992154.2996789

[6] Steptoe, A., Shankar, A., Demakakos, P., & Wardle, J. (2013). Social isolation, loneliness, and all-cause mortality in older men and women. In Proceedings of the National Academy of Sciences of the United States of America (Vol. 110, pp. 5797– 5801).

[7] Clifford Nass and Youngme Moon. 2000. Machines and mindlessness: Social responses to computers. J. Soc. Issues 56, 1 (Spring 2000), 81-103. Retrieved from http://ldt.stanford.edu/~ejbailey/02_FALL/ED_

[8] Irene Lopatovska and Harriet Williams. 2018. Personification of the Amazon Alexa: BFF or a Mindless Companion. In Proceedings of the 2018 Conference on Human Information Interaction & Retrieval (CHIIR '18). ACM, New York, NY, USA, 265-268. DOI: https://doi.org/10.1145/3176349.3176868

[9] Rock Leung, Charlotte Tang, Shathel Haddad, Joanna Mcgrenere, Peter Graf, and Vilia Ingriany. 2012. How Older Adults Learn to Use Mobile Devices: Survey and Field Investigations. ACM Trans. Access. Comput. 4, 3, Article 11 (December 2012), 33 pages. DOI: https://doi.org/10.1145/2399193.2399195